\documentclass[]{IEEEphot}

\jvol{xx}
\jnum{xx}
\jmonth{November}
\pubyear{2020}

\usepackage{enumerate}
\usepackage{multirow}
\usepackage{booktabs}
\usepackage{threeparttable}
\usepackage{graphicx}
\usepackage{cite}
\usepackage{amsfonts,amssymb}

\begin{document}

\title{Rotation Based Slice Error Correction Protocol for Continuous-variable Quantum Key Distribution and its Implementation with Polar Codes}

\author{Xuan Wen $^{1}$, Qiong Li $^{1,*}$, Haokun Mao $^{1}$, Xiaojun Wen $^{2,*}$, Nan Chen$^{3}$}
\affil{$^{1}$ Department of Computer Science and Technology, Harbin Institute of Technology, Harbin, China\\
$^{2}$ Shenzhen Polytechnic, Shenzhen, Guangdong, China\\
$^{3}$ School of Foreign Languages, Harbin Institute of Technology, Harbin, China
}

\maketitle

\begin{receivedinfo}%
Corresponding author: Qiong Li and Xiaojun Wen (e-mail: qiongli@hit.edu.cn; wxjun@szpt.edu.cn)
\end{receivedinfo}

\begin{abstract}
Reconciliation is an essential procedure for continuous-variable quantum key distribution (CV-QKD). As the most commonly used reconciliation protocol in short-distance CV-QKD, the slice error correction (SEC) allows a system to distill more than 1 bit from each pulse. However, its quantization efficiency is greatly affected by the noisy channel with a low signal-to-noise ratio (SNR), which usually limits the secure distance to about 30 km. In this paper, an improved SEC protocol, named Rotation-based SEC (RSEC), is proposed through performing a random orthogonal rotation on the raw data before quantization, and deducing a new estimator for quantized sequences. Moreover, the RSEC protocol is implemented with polar codes. Experimental results show that the proposed protocol can reach up to a quantization efficiency of about 99\%, and maintains at around 96\% even at the relatively low SNRs $(0.5,1)$, which theoretically extends the secure distance to about 45 km. When implemented with the polar codes with block length of 16 Mb, the RSEC can achieve a reconciliation efficiency of above 95\%, which outperforms all previous SEC schemes. In terms of finite-size effects, we achieve a secret key rate of $7.83\times10^{-3}$ bits/pulse at a distance of 33.93 km (the corresponding SNR value is 1). These results indicate that the proposed protocol significantly improves the performance of SEC and is a competitive reconciliation scheme for the CV-QKD system.
\end{abstract}

\begin{IEEEkeywords}
Continuous-variable quantum key distribution, reconciliation, slice error correction, polar codes, finite-size effect.
\end{IEEEkeywords}

\section{Introduction}

\label{intro}
 Quantum key distribution (QKD), which enables two remote legitimate parties, Alice and Bob, to share information-theoretic secret keys against a potential eavesdropper, is a major practical quantum cryptography technology in quantum information\cite{1}. There are mainly two categories of QKD protocols, namely discrete-variable (DV) protocol \cite{2,3,4,5,6} and continuous-variable (CV) protocol \cite{7,8,9,10,11}, which respectively encode information on discrete variables (such as the polarization or the phase of single photons) and continuous variables (such as the quadratures of coherent states). The DV-QKD needs a high-cost single-photon detector requiring cryogenic temperatures to measure the received quantum state, which presents a challenge for its widespread implementation. Compared to the DV-QKD, the CV-QKD takes the advantage of using a standard and cost-effective detector that is routinely deployed in standard telecom components working at room temperature. The security proof of CV-QKD against general attacks has been provided\cite{12,13,14,15}. Moreover, many experiments of CV-QKD have been successfully implemented, especially the integrated silicon photonic chip for CV-QKD that offers new possibilities for the low-cost and portable quantum communication\cite{16}.

\begin{figure}[t]
\centering
\includegraphics[width= 0.95\textwidth]{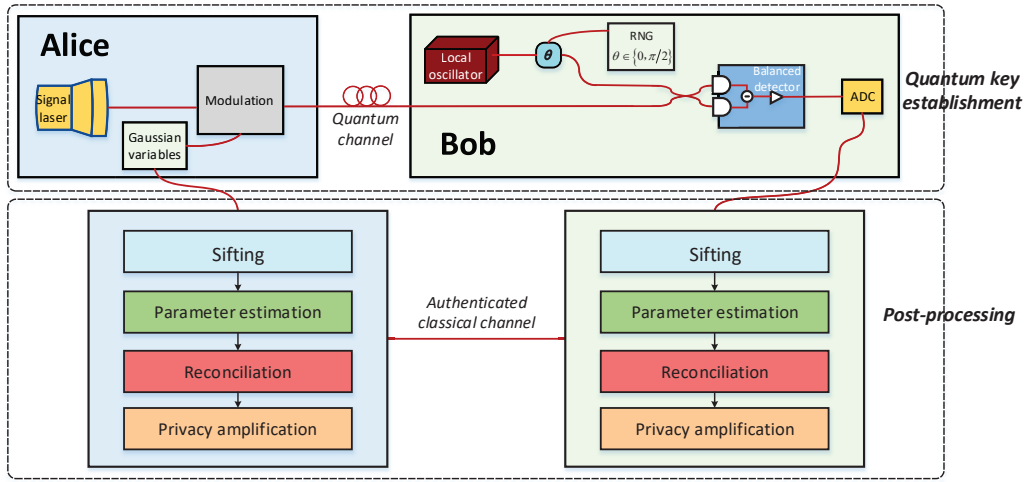}
\caption{Schematic diagram of the CV-QKD system}
\label{fig:1}
\end{figure}

A CV-QKD system mainly includes two consecutive phases \cite{7,8,9}: the quantum key establishment phase and the classical post-processing phase, which are illustrated in Fig.~\ref{fig:1}. In the first phase, Alice prepares a coherent state using two Gaussian variables and sends it to Bob through the quantum channel. Then Bob randomly chooses one of the two variables to measure his received coherent state and informs Alice of his choice.
Owing to the physical noises or the existence of Eve \cite{17} in a quantum channel, the raw data of the legitimate parties obtained from the first phase are weakly correlated and weakly secure continuous variables. To extract identical secret keys from their raw data, Alice and Bob subsequently perform a phase called post-processing including four main stages: sifting, parameter estimation \cite{18,19,20}, reconciliation \cite{21,22,23,24,25}, and privacy amplification \cite{26,27,28}. Reconciliation is a crucial stage for CV-QKD, which allows the legitimate parties to distill the corrected keys from their raw data via an authentic classical channel. Its performance affects the secret key rate and the secure distance of the practical CV-QKD system \cite{29,30,31,32}.

Up to now, various reconciliation schemes have been proposed for reconciling the raw data of CV-QKD. Originally, C. Silberhorn has proposed sign reconciliation that first quantifies the raw data to bit string by using the sign, and then corrects the error bits \cite{21}, but its low reconciliation efficiency limits its application.
 Subsequently, Cardinal et al have proposed SEC which chooses a set of quantization functions to convert a continuous variable into binary-value slices and then executes error correction on the quantized slices \cite{22,23}. Soon after, many researchers apply code-modulated techniques including multilevel coding (MLC) and multistage decoding (MSD) in SEC with  Low Density Parity Check (LDPC) codes to improve the reconciliation performance at high SNRs \cite{33,34}.
 The SEC scheme allows one to extract more than 1 bit of key from per pulse, especially at the high SNRs, but its quantization performance is poor at the low SNR of long-distance CV-QKD, which limits its secure distance to about 30 km.
 Afterward multidimensional reconciliation was proposed by Anthony Leverrier \cite{24}, which extends the secure distance from 30 km to above 50 km. Since the code rate of multidimensional reconciliation is limited to 1 bit per pulse, its related research is mainly focused on improving the reconciliation efficiency with LDPC codes, and especially with Multi-edge type LDPC (MET-LDPC) codes at low SNRs \cite{35,36,37,38,39,40,41}.

In summary, the existing research on reconciliation are mainly based on SEC and multidimensional reconciliation. These two schemes have their own advantages and disadvantages. Multidimensional reconciliation has a better quantization scenario than SEC reconciliation, so it can still achieve a high-efficiency reconciliation for the long-distance CV-QKD system with a noisy channel. However, its code rate is limited to 1 bit per pulse, which makes it more suitable for the long-distance CV-QKD system. Compared with multidimensional reconciliation, the SEC has advantages in extracting more than 1 bit of secret key per channel use. Limited by its quantization performance, SEC protocol is more suitable for the short-distance CV-QKD system.
As is known, the secret key rate of a QKD system will decrease rapidly with the increase of distance \cite{29}. Due to the technology immaturity of the physical device, the key generation rate of the long-distance CV-QKD system is generally low \cite{32,42}, which obviously cannot satisfy the communication demand. Therefore, to establish QKD networks \cite{43,44,45,46} with the short-distance QKD system is a practical scheme to provide relatively high-speed keys for secure communication at present \cite{47}.
In addition, the LDPC code is usually chosen to pursue a high reconciliation efficiency, but its matrix design is extremely difficult. By contrast, another common family of codes, polar codes, is relatively easier to construct and their recursive structure delivers excellent performance in practice.

In this research, our work focus on the improvement of SEC protocol and the reconciliation of the data with polar codes. The main contributions of this paper are as follows: (i) We improve the SEC protocol by first performing a random orthogonal rotation on the raw data before slice quantization and then providing a novel estimator for the quantized slices. Compared with SEC protocol, the improved protocol, named RSEC, has a higher quantization efficiency, which then increases the secret key rate and reconciliation efficiency; (ii) in order to accomplish the reconciliation of the correlated continuous variable in CV-QKD, we implement the RSEC protocol by combining the polar codes, achieving a high-efficiency reconciliation.

The rest of this paper is organized as follows: In Section \ref{sec:2}, the RSEC protocol is proposed to improve the SEC protocol. In Section \ref{sec:3}, the implementation of RSEC protocol with polar codes is described. In Section \ref{sec:4}, the experimental results and analysis of RSEC are given. Finally, the conclusion is drawn in Section \ref{sec:5}.

\section{Rotation based slice error correction (RSEC) protocol}
\label{sec:2}
 In this section, we briefly review the SEC reconciliation, and then put forward RSEC to improve the current SEC.
After the quantum key establishment phase of CV-QKD protocol, Alice and Bob share weakly correlated continuous-variable raw data due to the noises during the quantum transmission. The noises can safely be assumed to be Gaussian since it corresponds to the case of the optimal attack for Eve \cite{12}. Let $X = ({x_1},{x_2}, \cdots )$  and $Y = ({y_1},{y_2}, \cdots )$ corresponds to the correlated gaussian random variables of Alice and Bob respectively. Then, the correlated raw data can be modeled as $Y = X+Z$ with ${x_i} \sim N(0,\delta^2)$, ${z_i} \sim N(0,\sigma^2)$, where $Z=(z_1,z_2,\cdots)$, $\delta^2$ and $\sigma^2$ denote Alice's modulation variance and the noise variance respectively. In the direct reconciliation scenario, Alice's sequence is used as the target to correct Bob's sequence. On the contrary, the reverse reconciliation scenario uses Bob's sequence as the target to correct Alice's sequence. Generally, the latter scenario can obtain a higher secret key rate \cite{35,38}.
Without loss of generality, we only consider the reverse reconciliation in this research.
\subsection{Review of slice error correction}
In information reconciliation, Alice and Bob first perform an operation called quantization to convert the correlated values into binary sequences and then choose an error correction scheme to correct the binary sequence over an authenticated classical channel.
SEC is a generic reconciliation protocol \cite{22}. Its underlying idea is to convert Alice's and Bob's values into bit strings with slice function (i.e., quantization function), then apply an error correction scheme as a primitive, taking advantage of all available information to minimize the number of exchanged reconciliation messages.
 It works in two steps: First, Bob chooses a quantization function $\mathcal{S}(x):\mathbb{R} \to {\{ 0,1\}^m}$ to map his raw data to  $m$-slices binary digits, and informs Alice of the first $t$ slices (usually $t$ = 2 or 3), $\mathcal{S}(x)$ is a vector of slices $\mathcal{S}(x)=(\mathcal{S}_1(x),\cdots,\mathcal{S}_m(x))$; then, Bob sequentially deals with the remaining slice $k$ $(t+1 \le k \le m)$  by sending a syndrome of $\mathcal{S}_k(x)$
to Alice so that Alice can recover $\mathcal{S}_k(x)$ with a high probability.

In fact, the quantization function is to divide the set of real numbers $\mathbb{R}$ into $2^m$ intervals and then to assign different binary values to each of these intervals. There are two different schemes to construct the quantization function. The first construction scheme is to divide $\mathbb{R}$ with $2^m-1$ equidistant points. The second construction scheme freely chooses $2^m-1$ points to divide $\mathbb{R}$, which performs better but has a much higher computational complexity. The previous work has pointed out that the second scheme does not improve as much as the quantization efficiency compared with the first scheme \cite{33}. Therefore, we use the first scheme to construct the quantization function in this research.

In addition, previous studies have shown that the best bit assignment method is to assign the least significant bit of the binary representation of $a-1$ $(0 \le a-1 \le 2^m-1)$ to the first slice $\mathcal{S}_1(x)$ when $\tau_{a-1} \le x < \tau_a$ \cite{22}.
The variables $\tau_j$ divide the real numbers $\mathbb{R}$ into
$2^m$ intervals, where $1\le j \le 2^m-1$,  $\tau_0 = -\infty$, $\tau_{2^m} = +\infty$.
Then, each bit of $a-1$ is subsequently assigned up to the remaining slices. More specifically,
\begin{equation}
\label{Eq.1}
{\mathcal{S}_i}(x) = \left\{ \begin{array}{l}
0 \quad,{\rm{if}}~{\tau _{{2^i}n}} \le x < {\tau _{{2^i}n + {2^{i - 1}}}}\\
1 \quad,{\rm{otherwise}}
\end{array} \right.,
\end{equation}
where $1\le i\le m$ and $n$ is a nonnegative integer.

\subsection{Improving slice error correction with rotation}
In the decoding process of SEC, the slice sequences are corrected in sequence, hence the estimation of the current slice recursively depends on all previous slices. For this reason, the performance of SEC can be improved by reducing the BER $e_i$ of the previously decoded slices, $e_i$ denotes the probability that Alice makes a wrong estimate of Bob's slice value $\mathcal{S}_i(x)$. According to the characteristics of quantization function $\mathcal{S}(x)$, it is not hard to find that the last slice $\mathcal{S}_m(x)$ corresponds exactly to the sign of input variable $x$. Therefore, the quantization scheme of the last slice is similar to the multidimensional reconciliation which uses the sign of the rotated data as the target sequence.
As is known, multidimensional reconciliation usually performs better than the SEC reconciliation in estimating the quantized values, especially at a low SNR \cite{24}. For each slice, although obtained the first few slices, Alice still needs to infer infer Bob's slice value in a certain number of intervals. Taking the case of $m = 4$ slices as an example, if Alice gets the first two slices $(\mathcal{S}_1(x),\mathcal{S}_2(x))=(0,1)$, she needs to estimate Bob's slice $\mathcal{S}_3(x)$ among four intervals, i.e., $(\tau_2,\tau_3)$, $(\tau_6,\tau_7)$, $(\tau_{10},\tau_{11})$, $(\tau_{14},\tau_{15})$ to satisfy $(\mathcal{S}_1(x),\mathcal{S}_2(x))=(0,1)$. However, multidimensional reconciliation calculates the probabilities of Bob's quantized value with joint density function directly, which leads to more accurate estimations.

Consequently, to reduce the BER of the slice, we could execute a random orthogonal rotation on the raw data before the slice quantization and then infer the last slice $\mathcal{S}_m(x)$
according to multidimensional reconciliation. After decoding the $m$-th slice, Alice corrects the remaining slices in order. Assuming that Alice and Bob agree on the quantization function $\mathcal{S}(x)$ and the dimension $d$ of the orthogonal matrix, the procedure of our improved protocol for reverse reconciliation is shown in Fig.~\ref{fig:2}. The detailed process is described as follows:
 \begin{figure}[h]
\centering
\includegraphics[width= 0.85\textwidth]{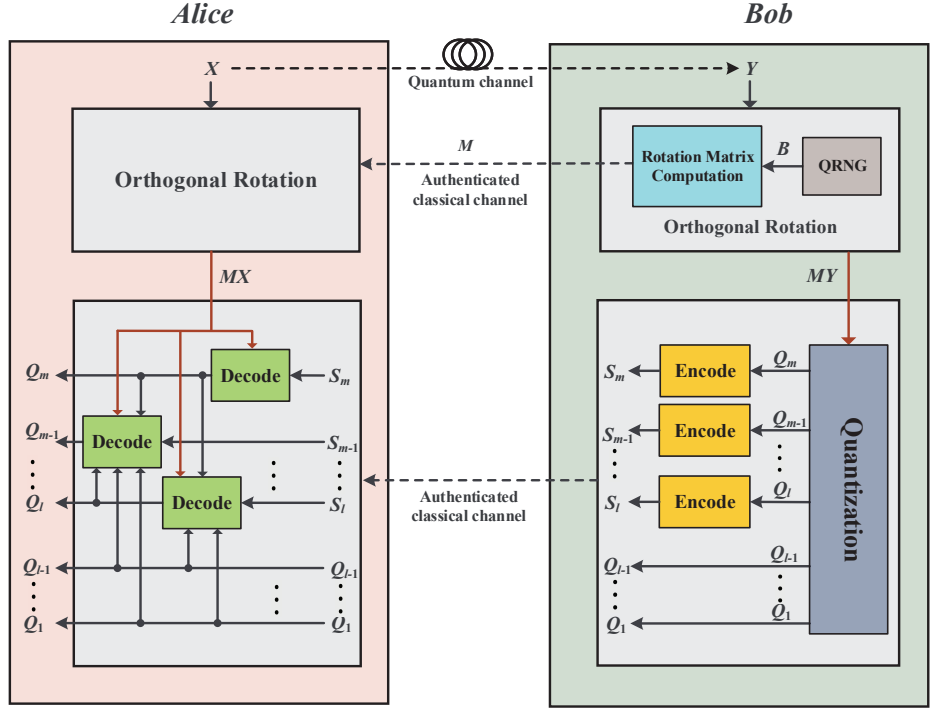}
\caption{Procedure of RSEC protocol for the continuous variables of CV-QKD system}
\label{fig:2}
\end{figure}

\textbf{\emph{Step 1}}: Alice and Bob divide their raw data into $d$-dimensional vectors as $X=\{x_i\}^d$, $Y=\{y_i\}^d$. Bob randomly generates a bit string $B=\{b_i\}^d$ and chooses a point $\mathcal{U}=\{\mu_i\}^d$ on unit sphere $\mathcal{O}^{d-1}$ adjacent to the point $\mathcal{U}_B =\{\frac{(- 1)^{b_i}}{\sqrt d}\}^d$. Then he calculates an orthogonal matrix $M$ satisfying $MT = \mathcal{U}$ for rotating $Y$ to $Y'=MY$, and informs Alice of the matrix $M$, where $T=\{t_i\}^d$, ${t_i} = \frac{{{y_i}}}{{\left\| Y \right\|}}$.

\textbf{\emph{Step 2}}: After receiving Bob's orthogonal matrix $M$, Alice performs the same rotation on $X$ and gets the rotated data $X'=MX$.

\textbf{\emph{Step 3}}: Bob quantizes his rotated data $Y'$ into $m$-slice bit vectors with the quantization function $\mathcal{S}(x)$ such as Eq.(\ref{Eq.1}), and sends the quantized slice values of the $1 \sim (l-1)$ slices $Q_1,\cdots,Q_{l-1}$ to Alice, where $Q_i=\mathcal{S}_i(Y')$.

\textbf{\emph{Step 4}}: Alice constructs a bit string $\hat Q_m$ of the $m$-th slice $Q_m$ from her rotated data $X'$ using the slice estimator $\hat {\mathcal{S}}$ as Eq.(\ref{Eq.13}) in Section \ref{sec:2.3}. Subsequently, Bob uses a chosen error correction codes to generate a syndrome $S_m$ so that Alice aligns her bit string $\hat Q_m$ on the sequence $Q_m$.

\textbf{\emph{Step 5}}: For each subsequent slice $k$, $l \le k < m$, Alice constructs a new
string by applying the slice estimator $\hat {\mathcal{S}}$ to $X'$, and taking into account the disclosed slices $Q_1,\cdots,Q_{l-1}$ and the previously corrected bit strings $Q_l,\cdots,Q_{k-1},$ $Q_m$. Again, Alice aligns her bit string to Bob's sequence $Q_k$ using their chosen error correction codes and corresponding syndrome $S_k$.

\subsection{Slice Estimator of RSEC}
\label{sec:2.3}
In the decoding stage of the RSEC reconciliation, we need to use the side information to estimate Bob's quantized slices first. Let us now detail the expressions we proposed. 
According to the decoding process, we first estimate the last slice $m$ of Bob. As is known,
\begin{equation}
\begin{array}{c}
Y' - X' = MY - MX\\
 = MZ.
\end{array}
\end{equation}
where $M = {({m_{ij}})_{d \times d}}$ is the rotation matrix, and $Z = \{z_i\}^d$ follows Gaussian distribution, ${z_i}\sim N{(0,{\sigma^2})}$. 

Because Gaussian variables have linear translation invariance, i.e., the linear combination of the independent Gaussian variables is still a Gaussian random variable, the random variable $Z' =Y' - X'$  has the same probability distribution as $Z$, i.e., $Y' - X'\sim N{(0,{\sigma ^2})^d}$. In addition, according to the characteristics of quantization function $\mathcal{S}(x)$, the bit string $Q_m=(Q^1_m,Q^2_m,\cdots)$ corresponds to the sign of the rotated data $Y'$, i.e., if $y'_i \ge 0$, $Q^i_m=1$, else, $Q^i_m=0,~i=1,2,\cdots$. Here, we use $Q^j_k$ to denote the $k$-th slice of $y'_j$. Hence, we obtain the conditional probability of ${Q^i_m}$ as follows
 \begin{equation}
{P_m}({Q^i_m}|{x'_i}) = \frac{K}{{\sqrt {2\pi {\sigma ^2}} }}{e^{\frac{{ - \left( {\mathcal{J}({Q^i_m})|y'_i| - {x'_i}} \right)^2}}{{2{\sigma ^2}}}}},
 \end{equation}
where $\mathcal{J}({x}) = {( - 1)^{{x} + 1}}$, $K$ is the normalization factor ${P_m}({Q^i_m} = 0|{x_i}) + {P_m}({Q^i_m} = 1|{x_i}) = 1$. By integrating the conditional probability into a parameter, we get the soft information called log likelihood ratio (LLR), which is a very useful parameter for estimation, as follows
 \begin{equation}
 \label{Eq.8}
 \ln \frac{{{P_m}({Q^i_m}=0|{x'_i})}}{{{P_m}({Q^i_m}=1|{x'_i})}} =  - \frac{{2{x'_i}|{y'_i}|}}{{{\sigma ^2}}}.
 \end{equation}

Given the transformation characteristics of the orthogonal rotation process, it is not difficult to deduce ${y'_i} = {\mu_i}||Y||$. If estimating $Q_m$ with Eq.(\ref{Eq.8}), Alice needs Bob to send his norm information $||Y||$, which will lead to heavy communication traffic and storage resources requirement. Fortunately, we have proposed a method that calculates the LLR without using the norm information of encoder in our previous work \cite{40}. Therefore, our protocol uses this improved method to calculate the LLR of $Q_m$ as follows
 \begin{equation}
 \label{Eq.9}
 {LLR}({Q^i_m}) = S_{nr}||X||\ln \frac{{1 - v({x'_i})}}{{1 + v({x'_i})}},
 \end{equation}
where $S_{nr}$ is the SNR of the quantum channel, and $v({x'_i})=\frac{{{x'_i}}}{{||X||}}$.

For the remaining slices $k$ $(l\le k < m)$, we derive their LLR with the corrected slices and the received $l-1$ slices as prior information. According to the previous analysis, we get the joint density function of the rotated data $X'$ and $Y'$ as Eq.(\ref{Eq.10}). Hence, the random variables $X'$ and $Y'$ follow the joint density function,
\begin{equation}
\label{Eq.10}
{f_{X'Y'}}(x,y) = \frac{1}{{2\pi \delta \sigma }}{e^{ - {{{{x}^2}} \mathord{\left/
 {\vphantom {{{{x}^2}} {2{\delta ^2}}}} \right.
 \kern-\nulldelimiterspace} {2{\delta ^2}}}}}{e^{ - {{{{(y - x)}^2}} \mathord{\left/
 {\vphantom {{{{(y - x)}^2}} {2{\sigma ^2}}}} \right.
 \kern-\nulldelimiterspace} {2{\sigma ^2}}}}}.
\end{equation}

According to Eq.(\ref{Eq.10}) and the characteristics of quantization function Eq.(\ref{Eq.1}), we derive that the conditional probability of $Q^i_k$ is expressed as
\begin{equation}
{P_k}(Q^i_k = b|{Q^i_{1, \cdots ,k - 1,m}},x'_i) = \sum\limits_\tau  {\int_{{\tau _{a - 1}}}^{{\tau _a}} {{f_{X'Y'}}(x'_i,y)dy} },
\end{equation}
where $\tau$ represents those quantization intervals satisfying $\mathcal{S}_{1, \cdots ,k - 1}(y) = B$, ${\mathcal{S}_k}(y) = b$, i.e., $\tau  = \{ ({\tau _{a - 1}},{\tau _a})|\forall y \in ({\tau _{a - 1}},{\tau _a}),\mathcal{S}_{1, \cdots ,k - 1,m}(y) = B,{\mathcal{S}_k}(y) = b\}$,
$b=$ 0 or 1, $B=(Q^i_{1, \cdots ,k - 1},Q^i_m)$ denotes the disclosed and corrected slices.

Accordingly, we get the initial LLR of $Q^i_k$ as Eq.(\ref{Eq.12}) to preliminarily estimate the rotated results of the $k$-th slice,
\begin{equation}
\label{Eq.12}
{LLR}({Q^i_k}) = \ln \left( {\frac{{\sum\limits_{{\tau ^0}} {\int_{{\tau _{a - 1}}}^{{\tau _a}} {{e^{ - {{x'{_i^2}} \mathord{\left/
 {\vphantom {{x'{_i^2}} {2{\delta ^2}}}} \right.
 \kern-\nulldelimiterspace} {2{\delta ^2}}}}}{e^{ - {{{{(y - {x'_i})}^2}} \mathord{\left/
 {\vphantom {{{{(y - {x_i})}^2}} {2{\sigma ^2}}}} \right.
 \kern-\nulldelimiterspace} {2{\sigma ^2}}}}}dy} } }}{{\sum\limits_{{\tau ^1}} {\int_{{\tau _{a' - 1}}}^{{\tau _{a'}}} {{e^{ - {{x'{_i^2}} \mathord{\left/
 {\vphantom {{x'{_i^2}} {2{\delta ^2}}}} \right.
 \kern-\nulldelimiterspace} {2{\delta ^2}}}}}{e^{ - {{{{(y - {x'_i})}^2}} \mathord{\left/
 {\vphantom {{{{(y - {x'_i})}^2}} {2{\sigma ^2}}}} \right.
 \kern-\nulldelimiterspace} {2{\sigma ^2}}}}}dy} } }}} \right),~l\le k < m,
\end{equation}
where $\tau ^0$ represents the  quantization intervals that satisfy ${\mathcal{S}_{1, \cdots ,k - 1,m}}(y) = B,{\mathcal{S}_m}(y) = 0$, and $\tau ^1$ satisfies ${\mathcal{S}_{1, \cdots ,k - 1,m}}(y) = B,{\mathcal{S}_k}(y) = 1$, respectively.

Based on the derived LLRs of each slice, the estimator $\hat {\mathcal{S}}$ of our RSEC reconciliation is constructed as follows
\begin{equation}
\label{Eq.13}
\hat {\mathcal{S}}(Q_j^i) = \left\{
{\begin{array}{*{20}{l}}
&{0\quad ,{\rm{if }}~LLR(Q_j^i) > 0}\\
&{1\quad ,{\rm{otherwise}}}
\end{array}} \right..
\end{equation}
Then, Alice can use Eq.(\ref{Eq.13}) to construct an initial estimation $\hat{Q}_j^i=\hat {\mathcal{S}}(Q_j^i)$ for Bob's slice value $Q_j^i$.

\subsection{Reconciliation efficiency}
Let us now discuss the reconciliation efficiency of the proposed protocol, which is an important indicator for evaluating the performance of the reconciliation procedure. As is known, the random orthogonal rotation operation on raw data does not expose any information of the rotated results \cite{24}. According to the efficiency expression of the SEC protocol, the reconciliation efficiency $\beta$ of the RSEC protocol can be expressed as
\begin{equation}
\beta  = \frac{{H(\mathcal{S}(Y'))  - m + \sum\limits_{i = 1}^m {{R_i}} }}{{I(X,Y)}},
\end{equation}
where $I(X;Y) = \frac{1}{2}\log_2 (1 + {S_{nr}})$ is the classical capacity of the quantum channel for Gaussian variables, $m$ denotes the number of slices of quantization function, and $R_i$ represents the code rate of the error correction scheme of the $i$-th slice. $H(\mathcal{S}(Y')) $ is the entropy of the slice sequences $\mathcal{S}(Y')$ which can be calculated as follows
\begin{equation}
H(\mathcal{S}(Y')) =  - \sum\limits_a {{P_a}\log_2 {P_a}},
\end{equation}
with
\begin{equation}
{P_a} = \frac{1}{2}\left( {erf\left( {\frac{{{\tau _a}}}{{\sqrt {2({\delta ^2} + {\sigma ^2})} }}} \right) - erf\left( {\frac{{{\tau _{a - 1}}}}{{\sqrt {2({\delta ^2} + {\sigma ^2})} }}} \right)} \right),
\end{equation}
where $\tau_a$ denotes the point dividing the real numbers $\mathbb{R}$, $1 \le a \le 2^m$, and $\tau_0 = -\infty$, $\tau_{2^m} = +\infty$. $\delta^2$ and $\sigma^2$ represent Alice's modulation variance and the noise variance respectively.

Generally, the code rate of the first $l-1$ slices are equal to $0$ since they are disclosed via the authentic classical channel.

\section{Implementation of RSEC with polar codes}
\label{sec:3}
After quantizing the continuous variables into strings of bits with slice functions, the legitimate parties are needed to further apply a classical error correction code to complete the reconciliation of the correlated raw data. In this section, we will implement the RSEC protocol with polar codes to distill the correct keys from the correlating raw data.

\subsection{Review of polar codes}
The polar code is an error correction code that has been strictly proven to achieve the Shannon capacity \cite{48}. The recursive structure of its encoding and decoding gives them good practical performance. What's more, it is relatively easier to construct than another commonly used code, i.e., LDPC code. Therefore, we choose polar codes to implement the RSEC protocol in this research. It should be noted that the RSEC can also be implemented with other error correction codes. Now, we briefly review the encoding and decoding of polar codes in traditional communication.

\subsubsection{Encoding}
The central idea of polar codes is to convert the $N$ individual copies of the channel $W$ into two different types of channels, i.e., error-free channel and completely noisy channel, through an operation called channel polarization --- channel combining and channel splitting. The information sender chooses the positions corresponding to the error-free channel to place her message bits (called information bits), and usually sets the remaining positions corresponding to the completely noisy channel as 0 (called frozen bits). The information bits and frozen bits together form a sequence $u^N_1$ of $N$ bits. We use the notation $u_1^n = ({u_1}, \cdots ,{u_n})$ to denote a row vector of $n$ bits. The sender encodes the sequence $u^N_1$ to a codeword $x^N_1$ by
\begin{equation}
x_1^N = u_1^NG,
\end{equation}
where $G$ is the generator matrix and defined as $G = {F^{ \otimes \log_2 N}}B$, $F^{ \otimes n}$  means to perform the Kronecker product $n$ times on the matrix $F \buildrel \Delta \over = \left[ {\begin{array}{*{20}{c}}
1&0\\
1&1
\end{array}} \right]$, and $B$ is a permutation matrix for executing the bit-reversal operation \cite{48}. Getting the codeword $x_1^N$, the sender transmits it to the information receiver for decoding.

\subsubsection{Decoding}
After the codeword  $x_1^N$ is transmitted through the channel, the receiver obtains a sequence $y_1^N$  which is a noise version of $x_1^N$. Then, he uses SC or SCL decoding algorithms to correct the error bits among $y_1^N$  with the given frozen bits. We here describe the receiver's decoding process with a SC decoding algorithm \cite{48}:
\begin{enumerate}
 \item Initialize the received information $y^N_1$ with channel transition probability $W(y|x)$  as
     \begin{equation}
     \label{Eq.18}
     L_1^{(j)}({y_j}) = {\frac{{W({y_j}|0)}}{{W({y_j}|1)}}},~~j = 1,2, \cdots ,N.
     \end{equation}
 \item Calculate the likelihood ratio (LR) of $u_j$ with the decoding results $\hat u_1^{j - 1} = ({\hat u_1},{\hat u_2}, $ $\cdots ,{\hat u_{j - 1}})$ of the previous $j-1$ bits as follows
    \begin{equation}
    L_N^{(j)}\left( {{y_j},\hat u_1^{j - 1}} \right) =  {\frac{{W_N^{(j)}(y_1^N,\hat u_1^{j - 1}|{u_j} = 0)}}{{W_N^{(j)}(y_1^N,\hat u_1^{j - 1}|{u_j} = 1)}}},
    \end{equation}
    where
    \begin{equation}
    W_N^{(j)}\left( {y_1^N,u_1^{j - 1}|{u_j}} \right) \buildrel \Delta \over = \frac{1}{{{2^{N - 1}}}}\sum\limits_{u_{j + 1}^N \in {{\{ 0,1\} }^{N - j}}} {{W_N}\left( {y_1^N|u_1^N} \right)}, \end{equation}
    and
    \begin{equation}
    {W_N}\left( {y_1^N|u_1^N} \right) = {W^N}\left( {y_1^N|x_1^N = u_1^NG} \right) = \prod\limits_{i = 1}^N {W\left( {{y_i}|{x_i}} \right)}.
    \end{equation}
 \item Generate the decision ${\hat u_j}$ of $u_j$ as
    \begin{equation}
    {\hat u_j} = \left\{
    {\begin{array}{*{20}{l}}
    &{{u_j}\quad,{\rm{if}}~j \in \mathcal{A}}\\
    &{0~\quad~,{\rm{if}}~j \notin \mathcal{A}~{\rm{ and }}~L_N^{(j)}\left( {y_1^N,\hat u_1^{j - 1}} \right) \ge 1}\\
    &{1~\quad~,{\rm{if}}~j \notin \mathcal{A}~{\rm{ and }}~L_N^{(j)}\left( {y_1^N,\hat u_1^{j - 1}} \right) < 1}
    \end{array}}
    \right.,
    \end{equation}
    where $\mathcal{A}$ is the position set of the frozen bits.
\end{enumerate}

After getting the $j$-th bit by step (iii), the process returns to step (ii) to decode the $(j + 1)$-th bit.

\subsection{Implementation process}
The reconciliation mode of CV-QKD is different from the traditional communication. In the traditional communication, the codeword is mixed with noises during the reconciliation. However, in a CV-QKD system, the two parties have already shared inconsistent data before the post-processing phase, in other words, the noise in the codeword has appeared before the reconciliation. Therefore, in order to correct the slice sequences of RSEC, it is necessary to establish a virtual channel for Alice and Bob to deal with the noise.

The encoding of polar codes is reversible: Encoding an input sequence $x$ twice, one can recover this sequence, i.e., $xGG = x$. This property can be used to establish a virtual channel as: Bob encodes a slice sequence $x$ to another sequence $u=xG$, and then sends the bits $u_{\mathcal{A}}$ corresponding to the frozen indices to Alice. Since $uG = (xG)G = x$, the slice sequence $x$ can be regarded as a polar codeword, Alice's initial estimation $\hat{\mathcal{S}}(x)$ of Bob's slice value can be viewed as the received codeword, and $u_{\mathcal{A}}$ corresponds to the frozen bits shared by the two parties. Therefore, a virtual channel can be established by using the above method.
\begin{figure}[bhtp]
\centering
\includegraphics[width= 0.95\textwidth]{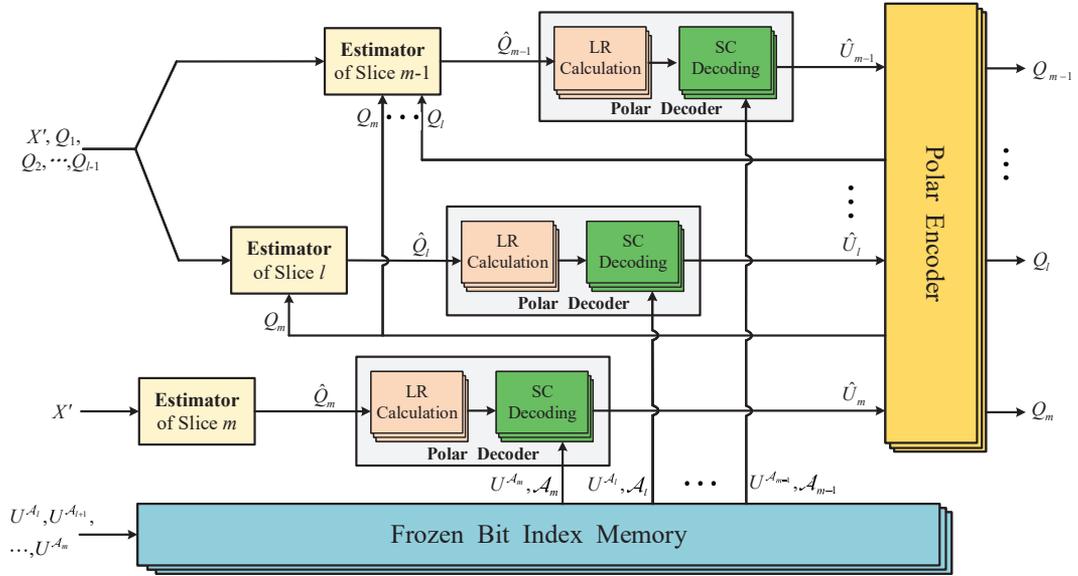}
\caption{Logic structure of the RSEC reconciliation with polar codes. The LR calculation modules provide the initial LR $L_{1,i}^{(j)}(\hat q_{j})$, and the frozen bit index memory stores the frozen bit $U^{\mathcal{A}_i}$ and frozen index set $\mathcal{A}_i$.}
\label{fig:3}
\end{figure}

Before launching the reconciliation with polar codes, Alice and Bob determine the code rate $R'_i$ of each slice according to the SNR and share the corresponding frozen index set $\mathcal{A}_i$. The frozen index set can be selected by a construction algorithm with consideration to $R'_i$. Then, the logic structure of the RSEC reconciliation with polar codes is shown in Fig.~\ref{fig:3}, in which the detailed implementation process is described as follows:

\textbf{\emph{Step 1}}: Alice and Bob convert their correlated data $X$, $Y$ to another continuous-variable sequence noted as $X'$, $Y'$ with random orthogonal rotation according to RSEC. Bob then quantizes $Y'$ into $m$ slice sequences $Q_1,Q_2,\cdots,Q_m$ with slice function and sends the first $l-1$ slices $Q_1,Q_2,\cdots,Q_{l-1}$ to Alice. Afterwards, they begin to reconcile the remaining slice sequences with polar codes in the order of $m,~l,~l+1,\cdots,~m-1$ slice.

\textbf{\emph{Step 2}}: Alice uses the proposed estimator Eq.(\ref{Eq.13}) to construct a bit string $\hat Q_i$ corresponding to Bob's slice sequence $Q_i$. Meanwhile, Bob encodes his slice sequence to $U = {Q_i}{G}$, and sends the bits $U^{\mathcal{A}_i}$ at the frozen positions to Alice;

\textbf{\emph{Step 3}}: Alice calculates the initial LR $L_{1,i}^{(j)}(\hat q_{j})$ as Eq.(\ref{Eq.23}), $j=1,2,\cdots,N$, and then makes a decision $\hat U$ on $U$ after getting the final LR $L_{N,i}^{(j)}( {\hat q_1^{N},{\hat u}_1^{j - 1}})$ in Eq.(\ref{Eq.24}). Afterwards, she can recover Bob's sequence $Q_i$ with a high probability by executing an encoding operation on $\hat U$.
 \begin{equation}
 \label{Eq.23}
 L_{1,i}^{(j)}({\hat q_{j}}) = {\frac{{W({\hat q_{j}}|0)}}{{W({\hat q_{j}}|1)}}},~~j = 1,2, \cdots ,N,
 \end{equation}
where $L_{1,i}^{(j)}({\hat q_{j}})$ is the initial LR corresponding to the $j$-th bit of $U$, $\hat q_{j}$ is the $j$-th bit of $\hat Q_i$, $U=(u_1,\cdots,u_N)$, $\hat U=(\hat u_1,\cdots,\hat u_N)$, the channel transition probability can be calculated as: if $y = x$, $W(y|x) = 1-e_i$, if $y\neq x$, $W(y|x) = e_i$. The bit error rate $e_i$ can be estimated in the stage of parameter estimation by executing the quantization operation on the extra raw data.
\begin{equation}
\label{Eq.24}
L_{N,i}^{(j)}( {\hat q_1^{N},{\hat u}_1^{j - 1}})= {\frac{{W_N^{(j)}(\hat q_1^{N},{\hat u}_1^{j - 1}|0)}}{{W_N^{(j)}(x_1^{'N},{\hat u}_1^{j - 1}|1)}}}.
\end{equation}

\noindent Moreover, the Eq.(\ref{Eq.24}) can evolve in a recursive manner as

\noindent if $j$ is odd, i.e., $j=2k-1$, then
\begin{equation}
L_{N,i}^{(2k - 1)}(\hat q_1^{N},\hat u_1^{2k - 2}) = f\left( {L_{N/2,i}^{(k)}\left( {\hat q_1^{N/2},\hat u_{1,o}^{2k - 2} \oplus \hat u_{1,e}^{2k - 2}} \right),L_{N/2,i}^{(k)}\left( {\hat q_{N/2 + 1}^{N},\hat u_{1,e}^{2k - 2}} \right)} \right),
\end{equation}
if $j$ is even, i.e., $j=2k$, then
\begin{equation}
L_{N,i}^{(2k)}(\hat u_1^{N},\hat u_1^{2k - 1}) = g\left( {L_{N/2,i}^{(j)}\left( {\hat q_1^{N/2},\hat u_{1,o}^{2k - 2} \oplus \hat u_{1,e}^{2k - 2}} \right),L_{N/2,i}^{(k)}\left( {\hat q_{N/2 + 1}^{N},\hat u_{1,e}^{2k - 2}} \right),{\hat u_{2k - 1}}} \right),
\end{equation}
where $f(a,b) = \frac{a\cdot b+1}{a+b}$, $g(a,b,s) = a^{1-2s} \cdot b$, we use $x^b_{a,o}$ to denote the odd terms of $x^b_a$, and $x^b_{a,e}$ denotes the even terms of $x^b_a$.

As a matter of fact, Alice can also use LLR as the soft information of polar codes for decoding. In this case, the initial LLR is calculated according to Eq.(\ref{Eq.9}) and Eq.(\ref{Eq.12}).

In order to ensure that the equation $\hat UG = {Q_i}$ holds with a high probability, Alice and Bob need to perform a cyclic redundancy check (CRC) to verify the decoding result $\hat U$. If $\hat U$ fails to pass CRC check, Alice and Bob give up on this slice $Q_i$. It is noted here that even if the decoding result passes the CRC check, undetected error bits may still exist. However, this situation rarely occurs and can be overlooked.

Because the CRC values will leak the information about $Q_i$, it is necessary to discard them. Therefore, the code rate $R_i$ of each slice is calculated as follows
\begin{equation}
{R_i} = {R'_i} - \frac{{{n_{crc}}}}{N},
\end{equation}
where $n_{crc}$ is the length of the CRC values.

\section{Experiment results and analysis}
\label{sec:4}
To evaluate the performance of the RSEC protocol, a series of experiments are carried out to compare their performances, including the quantization efficiency, reconciliation efficiency, and the secret key rate.

\subsection{Quantization efficiency of RSEC}
The principle of quantization is to minimize the information loss so that $I(X';\mathcal{S}(Y'))$ can be made arbitrarily close to the initially shared information $I(X;Y)$. After quantization, the mutual information $I(X';\mathcal{S}(Y'))$ shared by Alice and Bob can be expressed as
\begin{equation}
\label{Eq.28}
I(X';\mathcal{S}(Y')) = H(\mathcal{S}(Y')) - \left[ H({\mathcal{S}_m}(Y')|X') + \sum\nolimits_{i = 1}^{m-1} {H({\mathcal{S}_i}(Y')|X',{\mathcal{S}_{1 \ldots i - 1,m}}(Y'))} \right],
\end{equation}
where $\mathcal{S}(Y')=(\mathcal{S}_1(Y'),\cdots,\mathcal{S}_m(Y'))$ are the slice values of Bob.

Because the conditional entropy of Eq.(\ref{Eq.28}) recursively depends on all previously estimated results, calculating $I(X';\mathcal{S}(Y'))$ is not a simple task. For this reason, it is common practice to replace the conditional entropy with $H(e_i)$ equivalently \cite{22}. Then, the goal of quantization is simply to minimize each $e_i$, of which $H(e_i)$ is an increasing function for $0 \le e_i < 0.5$, $e_i$ is the BER of $i$-th slice. Therefore, the quantization efficiency $\beta _s$ can be measured with the Eq.(\ref{Eq.29}) equivalently \cite{22},
\begin{equation}
\label{Eq.29}
{\beta _s} = \frac{{H(\mathcal{S}(Y'))  - {H_e}}}{{I(X;Y)}},
\end{equation}
with ${H_e} = \sum\nolimits_{i = 1}^m {H({e_i})}$, $H(e_i)=-e_ilog_2(e_i)-(1-e_i)log_2(1-e_i)$.
\begin{figure}[htbp]
\centering
\includegraphics[width= 0.95\textwidth]{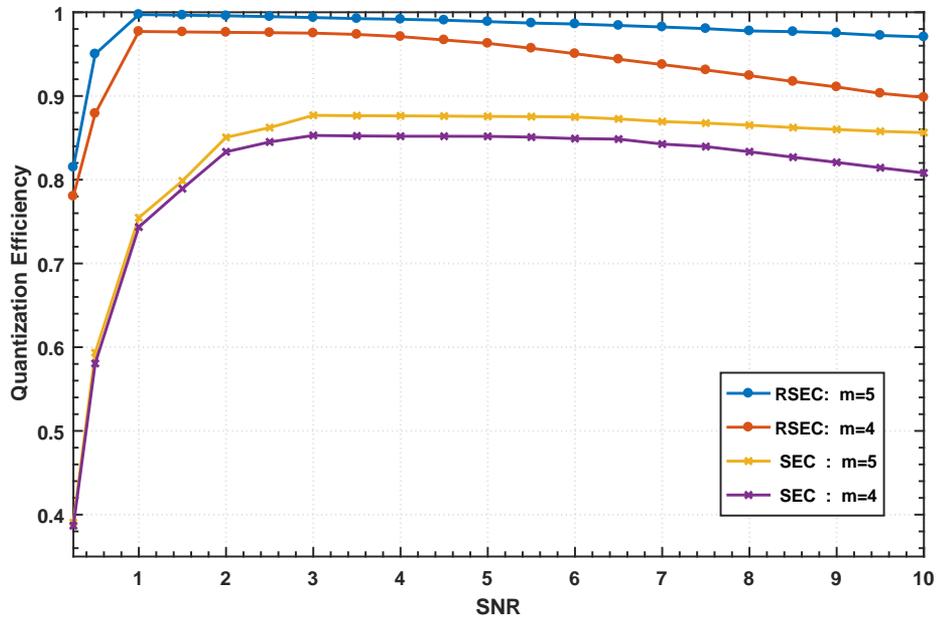}
\caption{Quantization efficiency of the RSEC and the SEC protocol with four and five slices at different SNR. The upper two curves denote the values of the RSEC protocol; the lower two curves correspond to the values of the SEC protocol.}
\label{fig:4}
\end{figure}

Figure \ref{fig:4} shows the quantization efficiency curves of SEC and RSEC at different SNR when $m = 4$ and $m = 5$. As can be seen from the figure, the quantization efficiency of SEC drops sharply for SNR $< 3$, which confirms that SEC reconciliation usually performs poorly at low SNRs. By contrast, RSEC can still maintain a high quantization efficiency $\beta _s>96\%$ almost for all SNRs $< 3$, and even achieve above 99\% quantization efficiency in the range of SNR $(1,3)$ when adopts the five-slice scheme.
The primary reason for the better performance of the proposed RSEC over the SEC protocol is attributed to our new estimator.
With the orthogonal rotation, our estimator can estimate Bob's slice sequences more accurately, especially for the slice that is decoded first, and thus the error rate $e_i$  decreases accordingly.
Moreover, the result in Fig.~\ref{fig:4} also confirms the following basic facts. For a fixed SNR, the higher the number of slices, the lower the information loss caused by quantization.

\subsection{Reconciliation efficiency of RSEC with polar codes}
It appears that the polarization speed of polar codes is highly dependent on the channel \cite{49}. Compared with the Binary Input Additive White Gaussian Noise Channel (BIAWGNC), constructing polar code for a Binary Symmetric Channel (BSC) is relatively uncomplicated and more common. Moreover, a BSC can be established between the two parties if Alice makes an initial estimation of Bob's slice sequences using LLR values. Accordingly, in our experiments, we construct the polar codes on a BSC, and calculate the initial LR as Eq.(\ref{Eq.23}) for decoding the slice sequences.
 \begin{figure}[htbp]
 \centering
  \includegraphics[width= 0.95\textwidth]{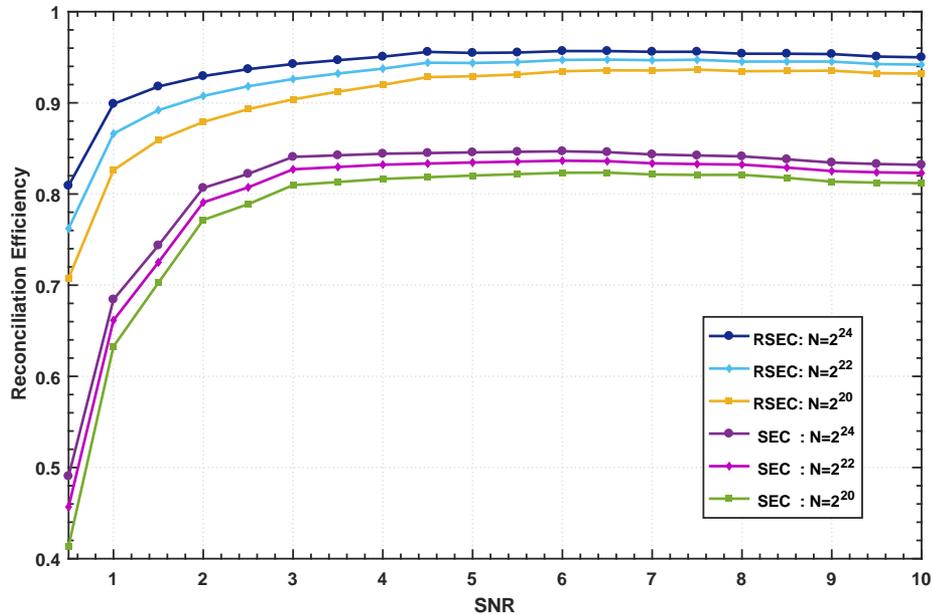}
\caption{Reconciliation efficiency of the RSEC and SEC protocol with polar codes at different SNR when the number of slices $m=5$. The upper three curves show the values of RSEC protocol; the lower three curves correspond to the values of SEC protocol. For RSEC and SEC protocol, their three curves from bottom to top represent the reconciliation efficiencies obtained with $N= 2^{20}$, $2^{22}$, $2^{24}$, respectively.}
\label{fig:5}
\end{figure}

Figure \ref{fig:5} compares the reconciliation efficiencies of the RSEC and the SEC protocol with polar codes  when $m=5$. The $32$-bit CRC is adopted for polar codes to check the decoding results, i.e., $n_{crc}=32$, and the eight-dimensional orthogonal matrix is used in rotation. For a fixed SNR value and different block length, 1000 blocks of raw data are generated to measure the reconciliation performance. The experimental results are obtained with FER $\le 0.1$, but a null BER in the blocks decoded successfully.

Combining Fig.~\ref{fig:4} and Fig.~\ref{fig:5}, it is not hard to find that the curvilinear trend of the quantization efficiency is basically consistent with that of the reconciliation efficiency, this is because the reconciliation scheme with good quantization performance usually performs better in reconciliation. Hence, the reconciliation efficiency of the proposed protocol is higher than the SEC protocol over the entire range in Fig.~\ref{fig:5} thanks to its higher quantization efficiency. As shown in Fig.~\ref{fig:5}, both the reconciliation efficiencies of RSEC and SEC increase with the increasing block length of polar codes since the decoding performance of polar codes will become better with the increase of its block size. The proposed RSEC protocol has an efficiency above 90\% over almost the entire range SNR $\ge 1$ for the block lengths starting from $2^{24}$, and even exceeds 95\% at SNR $>3$ where allows the system to distill more than 1 bit corrected key per raw data. It should be noted that RSEC has a high quantization efficiency in the SNR range (1,3) whereas its reconciliation efficiency is not so perfect. The reason is that the relatively low SNR leads to a high BER $>10\%$ in some noisy slices, and the decoding performance of polar codes decreases at high BER \cite{50}. In fact, the high quantization efficiency of RSEC allows the system to achieve a higher reconciliation efficiency by using a high-performance code.
\begin{table}[h]
	\centering
	\caption{Comparison of the reconciliation efficiencies between RSEC and some representative reconciliation works}
	\label{tab:1}
\begin{threeparttable}
\begin{tabular}{p{2.1cm}p{2.1cm}p{2.1cm}p{2.1cm}p{2.1cm}}
	\toprule[1pt]
    \multirow{2}*{SNR} &\multicolumn{4}{c}{Reconciliation efficiency $\beta$}\\[3pt]
    \cline{2-5}
    &Ref.~\cite{33}\tnote{a} &Ref.~\cite{23}\tnote{b} &Ref.~\cite{29}\tnote{c} & This work\\
    \midrule[0.8pt]
    3             & 94.1\%       & 79\%         & 88.7\%        & 94.85\%                   \\[3pt]
    5.12          & 94.4\%       & $-$          & $-$           & 95.53\%                   \\[3pt]
    7             & $-$          & 84\%         & $-$           & 95.60\%                   \\[3pt]
	14.57         & $95.8\%$     & $-$          & $-$           & 95.02\%                   \\[3pt]
    \bottomrule[1pt]
\end{tabular}
    \begin{tablenotes}
      \footnotesize
      \item[a] The slice number and error correction codes adopted in Ref.~\cite{33} are not reported in detail.
      \item[b] It implements the four-slice and five-slice SEC with the LDPC for blocks of $2\times10^5$.
      \item[c] It implements the four-slice SEC with the LDPC and BCH for blocks of $2\times10^5$ in a 25 km all-fiber CV-QKD system.
    \end{tablenotes}
\end{threeparttable}
\end{table}

In addition, we compare the reconciliation efficiency values with the representative works on SEC in Table \ref{tab:1}. As shown in the table, the proposed protocol almost improves all previously published reconciliation efficiencies in terms of the SEC protocol in the high SNR regime which is the main focus of the SEC reconciliation. In fact, the reconciliation efficiency values of Ref.~\cite{33} listed in the table are obtained under an optimistic situation of adopting the optimal number of slices and specially designed high-performance codes. Nevertheless, our reconciliation scheme still has a competitive advantage over Ref.~\cite{33} on the whole.
It should be noted that many achievements have also been made in multidimensional reconciliation, for example, Ref.~\cite{38} implements eight-dimensional reconciliation with $\beta =99\%$ and FER $=0.883$ using QC MET-LDPC code at SNR$=0.0283$, and Ref.~\cite{39} achieves $\beta = 93.40\%,~95.84\%,~96.99\%$ and FER $\le 0.375$ with eight-dimensional reconciliation based on MET-LDPC code at SNR $=0.160,~0.075,~0.029$, respectively. However, unlike the SEC protocol, the multidimensional reconciliation protocol is more suitable for the low SNRs rather than the high SNR regime. The existing works on multidimensional reconciliation are aimed at the extremely low SNRs and hardly provide the experimental results in the high SNR regime. Therefore, we mainly give a comparison with the representative results of the SEC protocol.

\subsection{Secret key rate of RSEC}
Assuming a collective Gaussian attack and accounting for the finite-size effects, the secret key rate of a CV-QKD system with reverse reconciliation can be expressed as \cite{29}:
\begin{equation}
K_{finite} = \frac{N_{data}}{N_{total}} \left[ \beta {I_{AB}} - {\chi _{BE}} - \Delta(N_{data}) \right],
\end{equation}
where $N_{total}$ is the total number of symbols sent from Alice to Bob, $N_{data}$ is the number of raw data used for key distillation, $\beta$ is the reconciliation efficiency, $I_{AB}$ denotes the mutual information between Alice and Bob, $\chi _{BE}$ denotes the Holevo bound on the information that Eve can obtain, and $\Delta(N_{data})$ is the finite-size offset factor.
$I_{AB}$ and $\chi _{BE}$ are related to the physical parameters including the transmittance $T$, the total noise $\chi_{total}$, and Alice's modulation variance $V_A$.
The transmittance $T$ of the quantum channel is defined as $T = {10^{ - {{\alpha \mathcal{L}} \mathord{\left/ {\vphantom {{\alpha d} {10}}} \right.\kern-\nulldelimiterspace} {10}}}}$, where $\alpha$ is the single-mode fiber transmission loss and $\mathcal{L}$ is the transmission distance. The total noise ${\chi _{{\rm{total}}}}$ consists of the channel added noise and the noise generated by Bob's detector, and be given by $ {\chi _{{\rm{total}}}} = {\chi _{{\rm{line}}}} + \frac{{{\chi _{\hom }}}}{T}$, where ${\chi _{{\rm{line}}}} = (\frac{1}{T} - 1) + \xi$ and ${\chi _{\hom }} = \frac{{1 + {V_{el}}}}{\eta } - 1$, $\xi$ is the excess channel noise, $V_{el}$ denotes the added electronic noise of Bob's detector, and $\eta$ represents the detector efficiency. The detailed calculation about $I_{AB}$, $\chi _{BE}$, and $\Delta(N_{data})$ can be found in Appendix A.

\begin{figure}[htbp]
\centering
 \includegraphics [width= 0.95\textwidth]{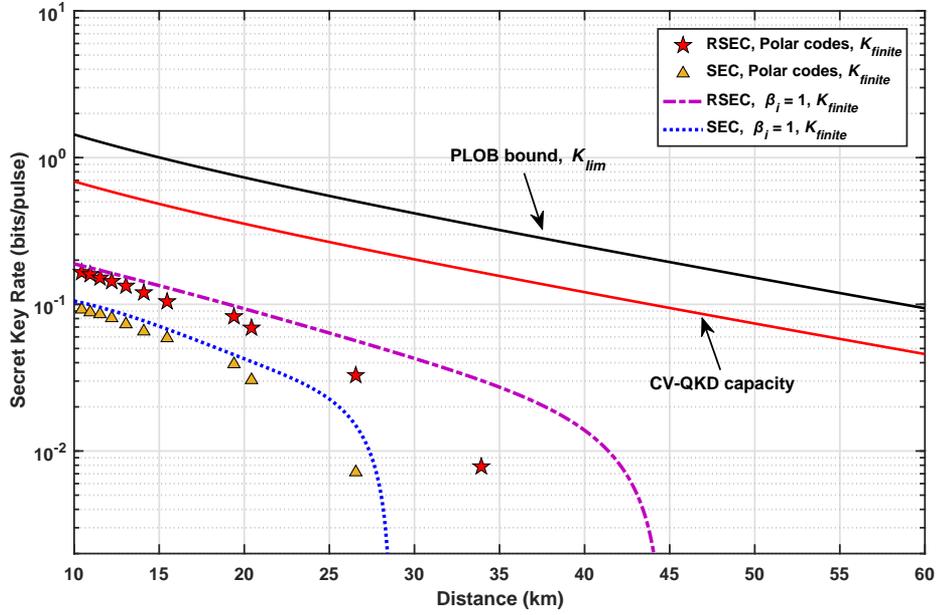}
 \caption{Finite secret key rate with $N_{total}=2^{40}$ vs. distance. The five-pointed stars correspond to the secret key rates using five-slice RSEC with polar codes of $N=2^{24}$; the triangle points correspond to the values using five-slice SEC with polar codes of $N=2^{24}$; the purple dot dash and blue dotted line represent the asymptotic theoretical secret key rates using five-slice RSEC and SEC with a perfect error correction scheme (i.e., $\beta_i=1$), respectively.
 Other parameters are as follows\cite{29}: $\alpha =0.2\rm{dB/km}$, $\xi=0.005$, $V_{el} = 0.041$, $\eta = 0.606$.}
 \label{fig:6}
\end{figure}

In our simulation, the experimental physical parameters reported in previously published work \cite{29} are used to characterize the CV-QKD system and quantum channel.
Optimizing the modulation variance $V_A$ for each transmission distance can maximize the SNR of a quantum channel.
The modulation variance $V_A$ in our work is adjusted according to the Ref.\cite{38}. Besides, we choose $N_{total} = 2N_{data}$ and the security parameter of $10^{-10}$ for $\Delta(N_{data})$\cite{18}.

Figure~\ref{fig:6} presents the finite secret key rates $K_{finite}$ over the  transmission distances with $N_{finite}=2^{40}$ bits. The five-pointed stars and triangle points compare the secret key rates achieved with polar codes of block lengths $N = 2^{24}$ bits, where the CV-QKD system using RSEC always provides higher secret key rates than that using SEC at the same transmission distance. Using RSEC reconciliation, we achieve a secret key rate of $7.83 \times 10^{-3}$ bits/pulse at a distance of $33.93$ km, while the CV-QKD system using SEC cannot provide any secret key.
In particular, assuming perfect error correction in the decoding of each slice, the RSEC protocol has more obvious advantages than the SEC protocol in the asymptotic secret key rate as the two dotted lines in Fig.~\ref{fig:6}. A perfect error correction scheme allows each slice to achieve its Shannon capacity, i.e., the efficiency $\beta_i$ of error correction in each slice is assumed as 1. Besides, with the increase of transmission distance, the secret key rate of CV-QKD decreases. This is because the SNR becomes lower with the increase of transmission distance, which leads to the reduction of quantification efficiency. Notably, when the transmission distance increases to about 30 km, the CV-QKD system using the SEC protocol can hardly generate any secret key. However, the RSEC protocol can theoretically extend the secure distance of the CV-QKD system to about 45 km.

There exists an upper bound called PLOB bound for the secret-key capacity of a lossy channel. The PLOB bound $K_{lim}$ is determined by the transmittance $T$ of channel and is given by\cite{51}
\begin{equation}
K_{lim} = -log_2(1-T).
\end{equation}
The black solid line in Fig.~\ref{fig:6} is the PLOB bound, which sets the fundamental rate limit for point-to-point QKD in the presence of loss. It is almost non-achievable for current protocols in the practical systems. Assuming the infinite-size keys and ideal conditions (such as unit detector efficiencies, zero dark count rates, zero intrinsic error, unit error correction efficiency, zero excess noise, etc.), the maximum rate of CV-QKD protocol (the red solid line) scales as $T/ln 4$, which is just $1/2$ of the PLOB bound \cite{51}. If taking the finite-size effect and the non-ideal factors of physical devices into account, the secret key rate of the practical CV-QKD systems will be much lower. As shown in Fig.~\ref{fig:6}, considering the non-ideal condition, the finite secret key rate of the CV-QKD system using RSEC can achieve $3.28\times 10^{-2}\sim1.652\times 10^{-1}$ bits/pulse at $10\sim27$ km, which is about $0.115$ of the PLOB bound. However, the system using SEC has a lower rate, which is just about $0.064$ of the PLOB bound, at $7.1\times 10^{-3}\sim9.22\times 10^{-2}$ bits/pulse.

The previous experimental results indicate that the proposed RSEC protocol is obviously advantageous. It significantly improves the quantization and reconciliation efficiency of SEC, which enables the CV-QKD system to achieve a higher secret key rate and a longer secure transmission distance. Overall, our work provides a better candidate for the application of the CV-QKD system.

\section{Conclusion}
\label{sec:5}
In this research, we analyzed the strategy of SEC protocol, and proposed modifications to improve its anti-noise ability by performing a random orthogonal rotation on the correlated raw data and deducing a slice estimator. The experimental comparisons of the original SEC protocol and the proposed RSEC protocol show that the modifications can reduce the information loss of the quantization and release the performance limitation of SEC at the relatively low SNR. Accordingly, both the secret key rate and the range of CV-QKD are increased. Moreover, in order to accomplish the reconciliation of the raw data in CV-QKD, we implemented the RSEC protocol by combing with the polar codes. The reconciliation efficiency of RSEC protocol can achieve above 95\% when the input scale adopts 16 Mb. Both theoretical and experimental analysis show that this work is a more suitable reconciliation scheme for the practical CV-QKD system.

\section*{Acknowledgements}
The authors wish to thank the anonymous reviewers for their valuable suggestions.
\section*{Appendix A}
\label{Appendix:a}
The mutual information between Alice and Bob $I_{AB}$ can be calculated by using Shannon's channel capacity \cite{22},
\begin{equation}
\label{AppendEq:1}
 I_{AB}=\frac {1}{2}log_2(1+S_{nr})=\frac {1}{2}log_2(\frac{V+\chi_{total}}{1+\chi_{total}}),
\end{equation}
where $V=V_A + 1$, $V_A$  represents Alice's modulation variance, and $\chi_{total}=\chi_{line}+\chi_{hom}/T$ represents the total noise between Alice and Bob as previously defined.

The Holevo bound on information available to Eve is given by
\begin{equation}
\label{AppendEq:2}
\chi_{BE}=G\left( \frac{\lambda_1 -1}{2} \right) + G\left( \frac{\lambda_2 -1}{2} \right) -
G\left( \frac{\lambda_3 -1}{2} \right) - G\left( \frac{\lambda_4 -1}{2} \right),
\end{equation}
where $G(x)=(x+1)log_2(x+1)-xlog_2(x)$, and the symplectic eigenvalues $\lambda_{1,2,3,4}$ are given by
\begin{equation}
\label{AppendEq:3}
\lambda_{1,2}^2=\frac{1}{2}(A\pm\sqrt{A^2-4B}),
\lambda_{3,4}^2=\frac{1}{2}(C\pm\sqrt{C^2-4D}),
\end{equation}
with
\begin{equation}
\label{AppendEq:4}
A=V^2(1-2T)+2T+T^2(V+\chi_{line})^2, B=T^2(V\chi_{line}+1)^2,
\end{equation}

\begin{equation}
\label{AppendEq:5}
C=\frac{V\sqrt{B}+T(V+\chi_{line})+A\chi_{hom}}{T(V+\chi_{total})}, D=\frac{V\sqrt{B}+B\chi_{hom}}{T(V+\chi_{total})}.
\end{equation}

When $N_{data} > 10^4$, the finite-size offset factor $\Delta(N_{data})$ can be approximated as follows\cite{12},
\begin{equation}
\label{AppendEq:8}
\Delta(N_{data})\approx 7\sqrt{\frac{log_2(2/{\epsilon})}{N_{data}}},
\end{equation}
where $\epsilon$ is the security parameter.

\bibliographystyle{IEEEtran}
\bibliography{mybibtex}



\end{document}